\documentclass{elsart}

\def\be{\begin{equation}}
\def\ee{\end{equation}}
\def\bea{\begin{eqnarray}}
\def\eea{\end{eqnarray}}

\usepackage{graphicx}
\usepackage{graphics}
\usepackage{amsmath}
\usepackage{hyperref}
\journal{Phyica A}
\begin{document}
\begin{frontmatter}

\title{Vortex transport and voltage noise in disordered superconductors}
\date{\today}
\author{Jayajit Das},
%\email{jayajit@vt.edu}
\author{Thomas J. Bullard},
%\email{tbullard@vt.edu}
and
\author{Uwe C. T\"auber}
%\email{tauber@vt.edu}
\address{Physics Department, Virginia Tech, Blacksburg, VA 24061-0435, USA}

\begin{abstract}
We study, by means of three-dimensional Monte Carlo simulations, the
current-voltage (IV) characteristics and the voltage noise spectrum at 
low temperatures of driven magnetic flux lines interacting with randomly 
placed point or columnar defects, as well as with periodically arranged 
linear pinning centers.
Near the depinning current $J_c$, the voltage noise spectrum $S(\omega)$
universally follows a $1/\omega^{\alpha}$ power law. For currents
$J > J_c$, distinct peaks appear in $S(\omega)$ which are considerably
more pronounced for extended as compared to point defects, and reflect
the spatial distribution of the correlated pinning centers.
\end{abstract}
\end{frontmatter}
%\section{Introduction}
The dynamics of flux lines in the mixed state of high-$T_c$ superconductors
has been a topic of intense interest in the past few years \cite{BLAT}. 
The area of research is motivated both by important technological questions 
and a quest of understanding complex non-equilibrium behavior of driven 
systems. 
Defects in the superconducting materials can either be localized (point 
pinning centers, such as oxygen vacancies) or extended. For example,
artificial columnar pins are introduced via high-energy ion irradiation
to pin the magnetic vortices in order to reduce energy loss due to dissipation 
when current is passed through the sample \cite{BLAT}. 
The non-equilibrium dynamics of flux lines in the presence of defects
generates non-trivial steady states which can be probed in experiments  
measuring, e.g., the current-voltage (IV) characteristics \cite{BLAT}, the 
voltage noise power spectrum \cite{MAE1,MAE2,SHOBHO1}, or local flux density
noise \cite{MAE2}.  
Analytical studies of the vortex dynamics are usually limited to asymptotic 
regimes \cite{BLAT,NEL,RISTE1} when the current $J$ is either much smaller 
or much larger than the critical current $J_c$. 
The intermediate current regime has been investigated by numerical techniques 
such as Langevin molecular dynamics simulations \cite{MARC,OLSON,TANG,REIC} or 
Monte Carlo techniques \cite{RYU1,RYU2}. 
These numerical simulations are largely restricted to two dimensions 
\cite{MARC,OLSON,TANG,REIC}, while very few three-dimensional studies have been
reported \cite{RYU1,RYU2,BLAT2}. Here we present a fully three-dimensional 
Monte Carlo simulation of vortex dynamics in the presence of both columnar and 
point defects. We focus on the effects of defect correlations in the voltage 
noise spectrum and the IV characteristics, some of which are inaccessible by 
two-dimensional simulations. 

We study the simplest possible model where the flux lines are non-interacting.
This may be viewed as a starting point towards a more realistic model that 
takes their mutual repulsion into account. At any rate, our present study 
should apply to a highly dilute vortex system. In spite of the simplistic 
non-interacting flux line model we find non-trivial signatures of broad-band 
noise (BBN) \cite{SHOBHO1,MAE1,MAE2} at $J\approx J_c$ as well as pronounced peaks in the 
voltage noise spectrum in the presence of extended disorder when $J>J_c$. 
We would like to emphasize that our results may apply quite generally for the 
dynamics of extended objects, such as charge density waves 
\cite{FISH1,FISH2}, interfaces \cite{FISH1,FISH2,RISTE2}, and polymers 
\cite{FISH1,FISH2}, driven through a disordered medium. 
Hence we consider a single magnetic vortex in the London approximation as an 
elastic line with line tension energy 
$E_L[{\bf r}(s)]=\frac{\varepsilon}{2}\int_{0}^{L} ds\, [ \Gamma^{-2} 
(d{\bf r}_{\perp}(s)/ds)^2+(dz(s)/ds)^2]$, 
where ${\bf r}(s) \equiv [{\bf r}_{\perp}(s),z(s)]$ describes the 
configuration of the flux line in three dimensions, and the crystalline axis 
${\bf c}$ of the superconducting material (as well as the external magnetic
field) is oriented parallel to $\hat{\bf z}$ (in a sample of thickness $L$). 
This linear elastic form of energy holds good as long as 
$|d{\bf r}_{\perp}(z)/dz| < 1/\Gamma$ \cite{BRAN}, where $\Gamma$ denotes the 
anisotropy ratio \cite{BLAT,NEL}. The line stiffness is given by 
$\varepsilon\approx \varepsilon_0 \ln(\lambda_{ab}/\xi_{ab})$ \cite{BLAT,NEL}, 
with $\varepsilon_0=(\phi_0/4\pi \lambda_{ab})^2$ 
($\phi_0 = h c / 2 e$ is the magnetic flux quantum). 
$\lambda_{ab}$ and $\xi_{ab}$ denote the penetration depth and the 
superconducting coherence length respectively in the $ab$ plane. 
For high-$T_c$ materials, $\Gamma \gg 1$ \cite{BLAT}, giving the quadratic 
form of $E_L[{\bf r}(s)]$ a wider range of validity. 
In the presence of (point or columnar) defects the flux line is described by
the free energy $F_L(T)$ defined through $\exp[-\beta F_L(T)L]=
\int D[{\bf r}(s)]\exp\left[-\beta E_L[{\bf r}(s)]-\beta\int_{0}^{L}ds 
V_p[{\bf r}(s)]\right]$, where $\beta=(k_BT)^{-1}$.  
We model the pinning potential as a sum of $N_p$ independent potential wells,
$V_p[{\bf r}(s)]=\sum_{k=1}^{N_p}U\Theta(r_p-|{\bf r}(s)-{\bf r}^{(p)}_k|)$,
where $r_p$ is the radius, $\Theta(x)$ denotes the Heaviside step function, 
and the ${\bf r}^{(p)}_k$ indicate the location of the pinning sites. 

In our Monte Carlo (MC) simulation each flux line is modeled by $N=L/a_0$ 
beads where the $i$th bead is located at ${\bf r}_i\equiv (x_i,y_i,z_i)$ and 
interacts with its nearest neighbors via a simple harmonic potential
$\frac{\varepsilon}{2}\sum_{\langle j \rangle = i-1,\, i+1}\,[\Gamma^{-2} 
({\bf r}_{\perp\,i}-{\bf r}_{\perp\,\langle j\rangle})^2+(|z_i-z_{\langle j 
\rangle}|-a_0)^2]$. 
The line is placed in a box of size $L_x \times L_y \times L_z$ with periodic 
boundary conditions in all directions.
The columnar pins and spherical pins are respectively modeled by constant 
cylindrical or spherical potential wells of (uniform) strength $U$ and radius 
$b_0$. The cylindrical wells are oriented parallel to the {\bf c} axis. 
We investigate the dynamics for three different distributions of defects, 
namely, (i) columnar pins distributed randomly or (ii) arranged in a square 
lattice in the $xy$ plane, and (iii) point pins distributed uniform randomly 
in the sample. When an external current ${\bf J}=-J\hat{\bf y}$ is applied it 
produces a Lorentz force 
${\bf f}_L=(1/c)\hat{z}\times {\bf J}=(J/c)\,\hat{\bf x}$, per unit length of 
the flux line. Since in the presence of weak currents ${\bf J}$ the flux line 
locally moves with an equilibrium dynamics one may incorporate the effect of 
the force in the MC by introducing an additional work term 
$-{\bf f}_L\cdot\int_{0}^{L}ds\, {\bf r}(s)$ in the energy $E_L[{\bf r}(s)]$ 
\cite{NEL,RYU1,RYU2}.
At $t=0$ the flux line starts off from a straight line configuration parallel 
to the $z$ axis. We have checked that the results in the steady state are 
independent of the initial configurations. At each trial a randomly chosen 
point on the line is updated according to a Metropolis algorithm \cite{BARK}. 
The point can then move in a random direction a maximum distance of 
$\triangle < b_0/\sqrt{3}$ to guarantee interaction with every defect. In the 
simulation the values $z_i$ are held fixed; we have checked \cite{BIG} that 
results do not change qualitatively even if these positions $z_i$ are allowed 
to fluctuate. 

The drift velocity of the flux line is proportional to the average velocity of 
the center of mass (CM) ${\bf v}_{cm}=\langle
\overline{[{\bf R}_{cm}(\tau)-{\bf R}_{cm}(0)]/\tau}\rangle$, where 
${\bf R}_{cm}(\tau)-{\bf R}_{cm}(0)$ is the distance traversed by the CM in 
a time interval of $\tau$; $\langle\cdot\cdot\cdot\rangle$ and the overbar 
denote the average over Monte Carlo steps (MCS) in the steady state and over 
different realizations of disorder, respectively. 
The voltage drop measured in experiments is caused by the induced electric 
field ${\bf E}={\bf B}\times {\bf v}_{cm}/c$ \cite{JOS}.
All length and energy scales are measured in units of $b_0$ and 
$\varepsilon_0$, respectively.
The average distance between the pins for a uniform random distribution or the 
lattice constant for the periodic array of columnar pins is taken as $d=15b_0$.
The parameters $\lambda_{ab}$, $\xi_{ab}$, $\varepsilon$, $U$, and $\Gamma$ are
chosen to be $56 b_0$, $0.64b_0$, $4\varepsilon_0$, $0.0075\varepsilon_0$, and 
$16$ respectively in the simulation; these numbers are consistent with the 
experimental data for high-$T_c$ materials \cite{BLAT}. 
Since here we are interested in studying the effects of defect correlations in 
the dynamics, we restrict our simulation to temperatures $T/T^{*}<1$, where 
$T^{*}=k_B^{-1}\sqrt{\varepsilon U}b_0$ is the temperature above which
entropic corrections due to thermal fluctuations become relevant for pinned 
flux lines \cite{NEL}.
Therefore thermally induced bending and wandering of the flux lines are largely
suppressed, and we can interpret our results in terms of low-temperature 
dynamics. In the simulation all the data have been collected in the the steady 
state where the system arrives at $t>10^5$ MCS and at $t>10^6$ MCS for columnar
and point pins, respectively. The size of the system ranges from 
$60\times 9 \times 60$ to $2000\times 9 \times 60$ in the simulations,
and the data are averaged over $20$ to $50$ realizations of disorder. The value
of $\tau$ ranges from $30$ to $500$ MCS in the simulation.

\begin{figure}
\includegraphics[scale=0.6]{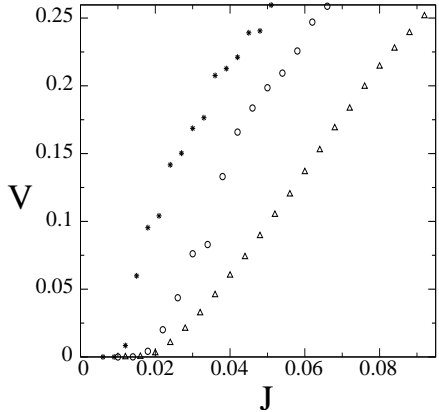}
\caption{\label{iv} IV characteristics for point pins ($\ast$) and columnar 
defects, both with random ($\circ$) and periodic ($\triangle$) spatial 
distribution. Note that $J_c$ is almost two times larger for columnar defects 
compared to point pins. For currents $J>J_c$ the randomly distributed columnar 
pins show a larger resistance than those with a periodic distribution 
(see text).}
\end{figure}

The IV characteristics displayed in Fig.~\ref{iv} were computed for random 
distributions of point and columnar pins, as well as for columnar defects 
arranged in a square lattice in the $xy$ plane at a temperature 
$T=0.25\times 10^{-3}<<T^{*}$.
The critical current $J_c$, which at zero temperature would indicate the
location of a non-equilibrium depinning transition, is higher for the columnar 
pins compared point disorder because of the increased defect correlations. 
The value of $J_c$ is the same for both random and periodic extended defect
distributions.
The flux line remains localized at few defect sites for $J\le J_c$ as 
thermal creep happens extremely rarely at $T<<T^{*}$ in the simulations. 
For currents $J>J_c$, the line becomes depinned and we observe that for 
columnar defects the ensuing voltage is lower for the periodic distribution as 
compared to the random distribution. 
This phenomenon can be understood in the following way. 
Suppose the energy of a flux line inside a columnar pin (say pin 1) of length 
$L$ is $U_1$ and the energy of the flux line in the neighboring pin (pin 2, at 
a distance $a$ from pin 1) where it arrives after a time $\Delta t(12)$ is 
$U_2$, then $U_2-U_1 \approx - Lf_La$, which is the effective energy 
barrier between $1$ and $2$ \cite{NEL}. 
The transit time is $\Delta t(12)\propto \exp(-\beta Lf_L a)$, and hence the 
mean velocity $v\propto a\exp(\beta L f_L a)$. 
Let us take a configuration where the three pins $A$, $B$ and $C$ are arranged 
in a straight line. The distance between $A$ and $B$ is $d_{AB}=a+\delta$ and 
the distance between $B$ and $C$ is $d_{BC}=a-\delta$. Note $\delta\neq 0$
and $\delta=0$ respectively for random and periodic distribution of the pins. 
Here the average velocity of hopping between neighboring pins is 
$\propto \frac{1}{2}[d_{AB}/\Delta t(AB)+d_{BC}/\Delta t(BC)]\propto 
v[\cosh(\beta L f_L\delta)+\delta \sinh(\beta L f_L\delta)]>v$. 
Therefore the average velocity of the flux line is 
less for a periodic defect arrangement as compared to a random distribution.
For periodic arrangement of the pins the IV characteristic depends strongly on 
the orientation of {\bf J} with respect to the lattice direction, for this 
determines the effective density of defects encountered by the moving vortex
\cite{BIG}.
\begin{figure}
\includegraphics[width=12.8cm,height=5.7cm]{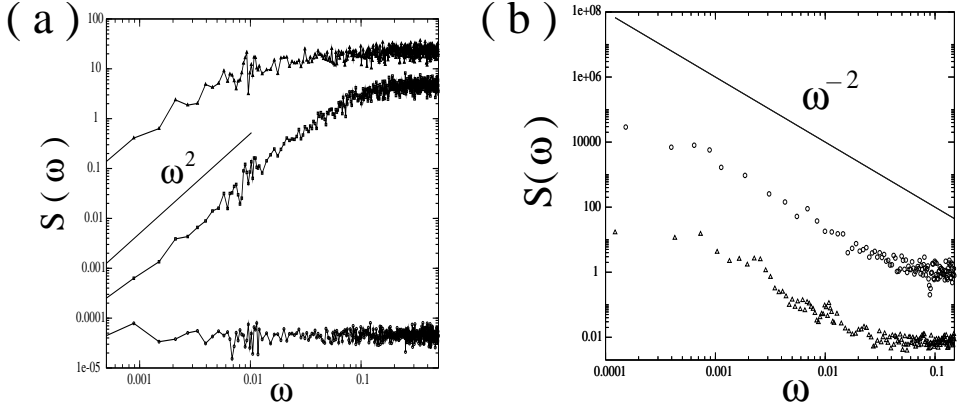}
\caption{\label{cnoi}
(a) Plot of the power spectrum $S(\omega)$ for $J=0$ for columnar pins at 
temperatures $T/T^{*}=6.5\times 10^{-4}$ (bottom), $6.5\times 10^{-2}$ 
(middle) and $T/T^{*}=3.3\times 10^{-1}$ (top). 
At $T/T^{*}=6.5\times 10^{-4}$, $r_g<<b_0$ and the voltage noise is essentially
thermal (white). At higher temperatures (for $T/T^{*}=6.5\times 10^{-2}$
and $T/T^{*}=3.3\times 10^{-1}$) the flux line starts to `feel' the potential 
well, rendering the fluctuations over large length scales correlated, as
reflected in the nonlinearity in the small $\omega$ behavior of $S(\omega)$. 
The nonlinearity follows a $\omega^2$ (solid line) behavior for small $\omega$ 
(see text). \newline
(b) Log-log plot of $S(\omega)$ vs. frequency $\omega$ for 
both columnar $(\triangle)$ and point($\circ$) defects near the depinning
threshold at $J\approx J_c$. The solid line, given as a guide, has slope $-2$.}
\end{figure}

The velocity (voltage) noise is calculated by computing the velocity 
fluctuations about $v_{cm}$. We evaluate the power spectrum 
$S(\omega)=\overline{\tilde{v}_x(\omega)\tilde{v}^{\ast}_x(\omega)}$, where 
$v_x(\omega)$ is the Fourier transform of the velocity fluctuation 
$\tilde{v}_x=v_x-\langle v_x \rangle$, with 
$v_x(t)=[X_{cm}(t+\tau)-X_{cm}(\tau)]/\tau$. The data are averaged over $50$ 
realizations of disorder for random distributions of defects.
At $T=0$ and for $J<J_c$, the flux line is bound to one or more pins depending 
on the type of defects present in the system, and $v_{cm}=0$. For temperatures 
close to $T=0$ the vortex is still trapped in the defect potential wells, and
$r_g^2=\frac{1}{N}\sum_{i=1}^{N}(r_{\perp i}-\bar{r}_{\perp})^2 << b_0^2$, 
where $\bar{r}_{\perp}=\frac{1}{N}\sum_{i=1}^{N}r_{\perp i}$. 
Therefore the line hardly tunnels out of the pins and $S(\omega)$ yields a 
white (thermal) noise spectrum (Fig.~\ref{cnoi} (a)).
At higher temperatures, when $r_g\approx b_0$, long length scale fluctuations 
can tunnel out of the defects and produce a nonlinear rise as 
$\propto \omega^2$ at small frequencies in $S(\omega)$ (Fig.\,\ref{cnoi}(a)).
This is because $S(\omega)$ is an even function, and $S(\omega=0)\approx 0$ 
(the equality is valid at $T=0$ and in the $L\rightarrow \infty$ limit). 
The non-white nonlinear part of $S(\omega)$ persists upto larger frequencies
for columnar pins as compared to the point defects because of increased 
correlations induced in the motion of the flux line by the correlated disorder
\cite{BIG}.
At temperatures $T > T^{*}$, $r_g>>b_0$ and the velocity fluctuations are 
dominated by thermal fluctuations; thus the high-frequency white-noise part of
$S(\omega)$ increasingly dominates the spectrum (Fig. \,\ref{cnoi}(a)).

At $J\approx J_c$, $S(\omega)$ shows a $1/\omega^{\alpha}$ behavior 
(Fig. \,\ref{cnoi}(b)), where $\alpha\approx \alpha_{MF}=2$ for almost a decade for 
both point and columnar defects with deviations at small and large frequencies.
This is to be interpreted as a remnant of the zero-temperature depinning 
transition, representing a non-equilibrium critical phenomenon 
\cite{ONU,KAR1,KAR2}. Its analysis through a functional renormalization group 
calculation gives $\alpha=1.5$ for point defects to one-loop order in three
dimensions \cite{KAR1,KAR2}, with corresponding mean-field value 
$\alpha_{MF}=2$.
Thus, while our data for lowest frequencies are affected by our finite system 
size, the $\omega^{-2}$ decay represents the finite-temperature mean-field like
behavior to be expected off criticality. The effective exponent $\alpha$ then
crosses over to smaller values, perhaps even its true critical value, at larger
frequencies. In experiments \cite{SHOBHO1}, the broad-band noise (BBN) features show a 
$\omega^{-\alpha}$ decay with $\alpha\approx 2$ which may similarly be a 
manifestation of the single-line depinning transition at a finite temperature. 

\begin{figure}
%\centerline{\epsfig{figure=./noise_per.eps,width=7.cm,height=6.5cm}}
\includegraphics[width=10cm,height=9cm]{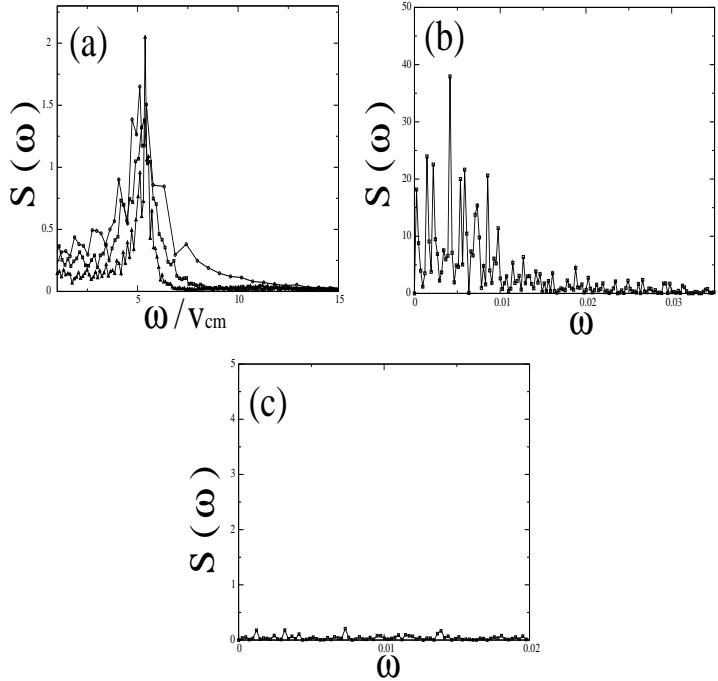}
\caption{\label{pernoi} (a) Plot of $S(\omega)$ vs. $\omega/v_{cm}$ for 
columnar pins arranged in a periodic array for $J=0.05 (\circ), \, J=0.07$ 
(square), and $J=0.09 (\triangle)$. The peaks occur at the same scaled 
frequency $\tilde{\omega}\propto v_{cm}/d$. Note that the width of the peak 
decreases as $J$ is increased indicating the wider separation between the time 
scales $\tau_r$ and $\tau_v$.
(b) Plot of $S(\omega)$ vs. $\omega$ for columnar pins distributed randomly in 
the sample for $J=0.05$ (square). 
Note the presence of secondary peaks in the spectrum (see text).
(c) Plot of $S(\omega)$ vs. $\omega$ for point pins at $J=0.05$ (square).
The pronounced peaks present in the case of columnar pins are considerably
suppressed (see text).}
\end{figure}

For currents $J>J_c$, two distinct time scales appear in the dynamics --- the 
residence time $\tau_r$ which is the duration of the vortices being trapped on 
the defects, and $\tau_v$, the time taken by the flux line to travel between
the pins. In the regime where these time scales are well separated and 
$\tau_r<\tau_v$, we can expect $\tau_v$ to produce pronounced peaks in the 
velocity noise spectrum. 
In Fig. \ref{pernoi}(a) we plot $S(\omega)$ at different driving currents
for the periodic distribution of columnar pins which reveals there is a single 
peak corresponding to the frequency $\tilde{\omega} \propto v_{cm}/d \propto 
1/\tau_v$. Note that the width of the peak decreases as the 
external current ${\bf J}$ increases, which widens the separation between 
$\tau_r$ and $\tau_v$. 
In a random defect distribution (Fig. \ref{pernoi}(b)) there would be a 
distribution of the time scale $\tau_v$ around the average value $d/v_{cm}$. 
Therefore there could be many secondary peaks present in the spectrum arising 
due to the distribution of $\tau_v$ in addition to the principle peak which 
corresponds to $\tau_r\propto d/v_{cm}$. The secondary peaks would eventually 
diminish as the distribution of the time scales will narrow down at $d/v_{cm}$ 
with more number of averaging on the defect configurations. 
Unlike for correlated defects, the time scales $\tau_r$ and $\tau_v$ are not 
well separated for point pins. The plot in Fig. \ref{pernoi}(c) reveals, as 
expected, that the peaks are considerably suppressed as compared to the 
situation for columnar defects (Fig. \ref{pernoi}(b)). 
This difference in $S(\omega)$ between columnar and point pins which results 
due to the lack of spatial correlation in the $z$ direction for the point 
disorder can be observed only in a three-dimensional dynamical simulation. 
It is due to a generic difference between correlated and localized pinning
potentials, and should survive even in the presence of interactions between the
flux lines.
The ensuing absence or presence of the narrow-band noise peaks in the voltage 
noise spectrum $S(\omega)$, and its specific features, can be utilized as a 
signature to identify and characterize the type of disorder present and 
responsible for flux pinning in experimental superconducting samples. 

In conclusion we have investigated the three-dimensional dynamics of 
non-interacting flux lines in the presence of both point and correlated 
defects. We find the nature of the current-voltage characteristics to depend
on the defect correlation and spatial distribution of the pinning centers. 
The voltage noise spectrum shows pronounced peaks in the case of correlated pins. 
The difference in the power spectrum at $J>J_c$ between columnar and point 
disorder can serve as a novel diagnostic tool to characterize the pinning
centers in superconducting samples. 

We would like to thank S. Bhattacharya, A. Maeda, B. Schmittmann, P. Sen, and
R. K. P. Zia for valuable discussions. 
This research has been supported by the National Science Foundation
(grant no. DMR 0075725) and the Jeffress Memorial Trust (grant no. J-594).


\begin{thebibliography}{00}
\bibitem{BLAT}
G. Blatter, M. V. Feigel'man, V. B. Geshkenbein, A. I. Larkin, and 
V. M. Vinokur, Rev. Mod. Phys. {\bf 66} (1994) 1125.

\bibitem{MAE1}
Y. Togawa, R. Abiru, K. Iwaya, H. Kitano, and A. Maeda, Phys. Rev. Lett. 
{\bf 85} (2000) 3716.

\bibitem{MAE2}
A. Maeda, T. Tsuboi, R. Abiru, Y. Togawa, H. Kitano, K. Iwaya, and 
T. Hanaguri, Phys. Rev. B {\bf 65} (2002) 054506.

\bibitem{SHOBHO1}
A. C. Marley, M. J. Higgins, and S. Bhattacharya, Phys. Rev. Lett. {\bf 74} 
(1995) 3029.

%\bibitem{SHOBHO2}
%R. D. Merithew, M. W. Rabin, M. B. Weissman, M. J. Higgins, and 
%S. Bhattacharya, Phys. Rev. Lett. {\bf 77} (1996) 3197.

\bibitem{NEL}
D. R. Nelson and V. M. Vinokur, Phys. Rev. B {\bf 48} (1993) 13060.

\bibitem{RISTE1}
D. R. Nelson, {\it Vortex line fluctuations in superconductors from elementary 
quantum mechanics}, in: Proceedings of the NATO Advanced Study Institute on 
{\it Phase Transitions and Relaxation in Systems with Competing Energy Scales},
eds. T. Riste and D. Sherrington (Kluwer Academic Publ., Boston, 1993), p. 95.

\bibitem{MARC}
M. Dong, M. C. Marchetti, A. A. Middleton, and V. M. Vinokur, Phys. Rev. Lett. 
{\bf 70} (1993) 662.

\bibitem{OLSON}
C. J. Olson, C. Reichhardt, and F. Nori, Phys. Rev. Lett. {\bf 80} (1998) 2197.

\bibitem{TANG}
C. Tang, S. Feng, and L. Golubovic, Phys. Rev. Lett. {\bf 72} (1994) 1264.

%\bibitem{CARN}
%G. Carneiro, Phys. Rev. B {\bf 62} (2000) R14661.

\bibitem{REIC}
C. Reichhardt and C. J. Olson, Phys. Rev. B {\bf 65} (2002) 0943011.

\bibitem{RYU1}
S. Ryu, A. Kapitulnik, and S. Doniach, Phys. Rev. Lett. {\bf 71} (1993) 4245. 

\bibitem{RYU2}
S. Ryu and D. Stroud, Phys. Rev. B {\bf 54} (1996) 1320.

\bibitem{BLAT2}
A. Sch\"onenberger, A. Larkin, E. Heeb, V. Geshkenbein, and G. Blatter, 
Phys. Rev. Lett. {\bf 77} (1996) 4636.

\bibitem{FISH1}
D. S. Fisher, Phys. Rep. {\bf 301} (1998) 113. 

\bibitem{FISH2}
M. Kardar, Phys. Rep. {\bf 301} (1998) 299.

\bibitem{RISTE2}
D. S. Fisher, {\it Low temperature phases, ordering and dynamics in random 
media}, in: Proceedings of the NATO Advanced Study Institute on {\it Phase 
Transitions and Relaxation in Systems with Competing Energy Scales}, eds. 
T. Riste and D. Sherrington (Kluwer Academic Publ., Boston, 1993), p. 1.

\bibitem{BRAN}
E. H. Brandt, Phys. Rev. Lett. {\bf 69} (1992) 1105.

%\bibitem{DAAN} 
%D. Frenkel and B. Smit, {\it Understanding Molecular Simulation: 
%From Algorithms to Applications}, 1st ed. (Academic Press, NY, 1996).

\bibitem{BARK}
M. E. J. Newman and G. Barkema, {\it Monte Carlo methods in Statistical 
Physics} (Clarendon Press, Oxford, 1989).

\bibitem{BIG} 
J. Das, T. J. Bullard, and U. C. T\"auber, in preparation.

\bibitem{JOS}
M. Tinkham, {\it Introduction to Superconductivity} (McGraw-Hill, NY, 1975).

\bibitem{ONU}
O. Narayan and D. S. Fisher, Phys. Rev. B {\bf 48} (1993) 7030.

\bibitem{KAR1}
D. Ertas and M. Kardar, Phys. Rev. Lett. {\bf 73} (1994) 1703. 

\bibitem{KAR2}
M. Kardar and D. Ertas, {\it Non-equilibrium dynamics of fluctuating lines}, 
in: Proceedings of the NATO Advanced Study Institute on {\it 
Scale Invariance, Interfaces, and Non-Equilibrium Dynamics}, eds. A. McKane, M. Droz, 
J. Vannimenus, and D. Wolf (Plenum, New York, 1995), p. 89.

\end{thebibliography}
\end{document}